\documentclass[10pt,a4paper]{article}

\def\heading #1{\bigbreak \begin{center} {\bf #1} \end{center}}

\title{A variational Monte Carlo calculation of dynamic
multipole polarizabilities and van der Waals coefficients of the PsH system}

\author{Massimo Mella \ \\
Dipartimento di Chimica Fisica ed Elettrochimica,\\ Universita' degli Studi
di Milano, via Golgi 19, 20133 Milano, Italy\\
Electronic mail: Massimo.Mella@unimi.it\\
\and 
Dario Bressanini$^a$, and Gabriele Morosi$^b$\\
Dipartimento di Scienze Chimiche, Fisiche e Matematiche,\\
Universita' dell'Insubria, \\
via Lucini 3, 22100 Como, Italy\\
$^a$ Electronic mail: Dario.Bressanini@uninsubria.it\\
$^b$ Electronic mail: Gabriele.Morosi@uninsubria.it\\}

\begin{document}

\maketitle

\bigbreak
\begin{abstract}
The first three dynamic multipole polarizabilities for the
ground state of hydrogen, helium, hydride ion, and positronium hydride PsH 
have been computed
using the variational Monte Carlo (VMC) method and explicitly 
correlated wave functions.
Results for the static dipole polarizability by means of the diffusion
Monte Carlo (DMC) method and the finite field approach show the
VMC results to be quite accurate. From these dynamic polarizabilities
van der Waals dispersion coefficients for the interaction of PsH with ordinary
electronic systems can be computed, allowing one
to predict the dispersion energy for the interaction
between PsH and less exotic atoms and molecules.
\end{abstract}

\subsubsection*{PACS number(s): 36.10.-k, 02.70.Lq}

\pagebreak

While experimentalists relay everyday on positrons and positronium
atoms (Ps) to collect information about microscopic features of
macroscopic systems like solutions, polymers and crystals,
much less effort has been devoted to the theoretical understanding
of the complex interactions that take place between ordinary
matter and positrons.
Among the explored avenues of this field, we mention the interest
in predicting the stability of classes of compounds
like e$^+$M and MPs [1-9],
~\nocite{ryz,yoshi1,maxcor,jiang,stras3,sil,clary,ho,frolov}
where M represents an atomic or molecular system,
and the calculation of the cross sections in the scattering
process of e$^+$ and Ps on a molecule or an atom [10-16].
~\nocite{mcalinden,sur,comi,ghosh,biswas,chaud,nath}

On the contrary,
the evaluation of the interaction energy between e$^+$M or MPs and 
a molecule or atom is an almost unexplored 
issue ~\cite{maxh2ps}. We believe this fact is primarily due to the 
need of a very accurate
trial wave function to describe correctly
the correlated motions of electrons and positrons.
So far, only variational calculations with explicitly correlated Gaussians (ECG)
~\cite{ryz,stras3} or Hylleraas-type functions ~\nocite{clary,ho,frolov}
[7-9], and the DMC method [2-4]
~\nocite{yoshi1,maxcor,jiang} 
have shown to be able to adequately recover the correlation
energy in positron-containing systems. 

Related to the calculation of the interaction energies is the calculation 
of second order properties
of positron-containing systems, a problem whose surface has been
barely scratched in the past ~\cite{sil}. These properties, specifically
the dynamic polarizabilities, are
strictly related to the van der Waals coefficients that describe the
long range interaction between systems ~\cite{szabo}, representing a
way to tackle the problem of the asymptotic intermolecular interactions.
Recently, Caffarel and Hess showed that these properties can be computed
by means of quantum Monte Carlo simulations ~\cite{caf1}
connecting the imaginary-time-dependent
dynamics of the unperturbed system with the transition probabilities of
a reference diffusion process. 
In this work we apply a modified version of their method to compute
dynamic multipole  polarizabilities for PsH, H, He, and H$^-$ as a way to
understand the behaviour of these systems when interacting with an
external field, and as a first step towards the definition of the interaction
potential between PsH and the ordinary matter.

As far as we know, the work by Le Sech and Silvi ~\cite{sil} is the only one
reporting calculations
on the effect of a constant electric field on PsH. In that work they
computed both the static dipole polarizability, 123 a.u., and the behaviour of 
the annihilation rate $\Gamma_{2\gamma}$ versus the intensity of
the field employing explicitly correlated wave functions, numerical
integration, and  a variation-perturbation approach. 
As by-product of our calculations of the potential energy curve of the 
e$^+$LiH system ~\cite{maxlihe+}, we obtained an estimation of the static dipole
polarizability of 49(2) a.u., a value quite different from
the one computed by Le Sech and Silvi. Since we believe this difference
to be too large to admit an explanation based on the different accuracy
of the methods used to compute this value, we plan to solve this
puzzle in this work.

In the method by Caffarel and Hess~\cite{caf1} the frequency-dependent 
second order correction to the ground state energy
is written as a sum of the two time-centered autocorrelation functions
of the perturbing potential $V$

\begin{equation}
E^{(2)}_{\pm}(\omega) = -\int _0 ^{\infty} e^{\pm t \omega} C_{VV}(t) \, dt
\label{pereq1}
\end{equation}
\noindent
where the autocorrelation function $C_{VV}(t)$ is given by

\begin{equation}
C_{VV}(t) = \langle V(0)V(t)\rangle_{\Psi_0^2} - 
\langle V(0)^2\rangle_{\Psi_0^2}
\label{pereq2}
\end{equation}

\noindent
Here, $\langle ...\rangle_{\Psi_0^2}$ 
indicates that the average has to be taken
using the Langevin dynamics that samples the square of the exact 
ground state wave function of the unperturbed system. 

Caffarel and Hess ~\cite{caf1} showed that it is possible to 
compute $C_{VV}(t)$ employing an optimized trial wave function and 
the pure-diffusion Monte Carlo
(PDMC) method, an alternative algorithm to the commonly used DMC with
branching, where each walker explicitly carries its own weight along all
the simulation ~\cite{reybook}. 

In their work on He and H$_2$, Caffarel, R\'erat, and Pouchan ~\cite{caf2}
reported that the autocorrelation function $C_{VV}(t)$ becomes
dominated by the noise at large times, and this fact might be due
to the fluctuations of the walker weights that increase during a PDMC simulation,
while the value of the autocorrelation function itself becomes smaller.
While the second effect is intrinsic to the stochastic method,
the first can be reduced employing a more accurate trial wave function
that is able to reduce the weight fluctuations.
Another possibility, giving up the exactness of the method (i.e. not sampling the exact $\Psi_0^2$), is represented by the sampling of a quite accurate
trial wave function without carrying around the weight for each walker,
a method we call Perturbation Theory variational Monte Carlo (PT-VMC).
This algorithm can be useful for those systems whose autocorrelation 
function has a large decaying time, as the case of H$^-$ and PsH. 
This large decaying time will increase the fluctuations of
the carried weights, and hence the statistical noise in the
autocorrelation functions in the long-time region.
 
As a test of the correctness of our computer program
and of the accuracy of the method, we computed
the first three autocorrelation functions, and hence 
the dynamic polarizabilities up to the octupolar one, 
for the two systems H and He. The analytical forms of the perturbing potentials
were taken from Ref. ~\cite{caf2}.
While for H we employed the exact ground
state wave function and compared with the analytic values of the 
multipole polarizability ~\cite{szabo}, for the He case we used a 25 term 
Hylleraas-type
wave function optimized by means of the standard energy minimization ~\cite{gab}.
We fitted the numerical $C_{VV}(t)$ results of our simulations 
with a linear combination of three exponential functions
 
\begin{equation}
\label{pereq2tris}
C_{VV}(t) \simeq \sum_{i=1}^{3} a_i e^{-\lambda_i t}
\end{equation}
\noindent
in order to have an analytical representation of the autocorrelation
functions at all the times. Since it is important to reproduce accurately
the long time behaviour of $C_{VV}$, the smallest $\lambda_i$
in Eq. \ref{pereq2tris} was independently calculated fitting 
$ln \left[ C_{VV}\right]$ in the long time region with a first order 
polynomial. This choice was found to improve sensibly
the goodness of the total fitting in this time range.

These analytical representations of $C_{VV}$ allow us to compute easily
the integrals in Eq. \ref{pereq1} and to obtain  simple expressions
of $\alpha (\omega)$.  
The parameters obtained by the fitting procedure
are available from the authors upon requests.

For both systems we found excellent agreement of the static
polarizabilities (H $\alpha_{dip}=$4.495 au, $\alpha_{quad}=$15.034 au,
$\alpha_{oct}=$133.105 au; He $\alpha_{dip}=$1.382 au, 
$\alpha_{quad}=$2.401 au, $\alpha_{oct}=$10.367 au) 
with the exact results for H ~\cite{szabo},
with PDMC results by Caffarel, R\'erat, and Pouchan ~\cite{caf2},
with Glover and Weinhold upper and lower bounds for He ~\cite{glo}, and
with the accurate results by Yan {\it et al.} ~\cite{drake}.

At this point we would like to stress that, although in the PT-VMC method
the walkers carry always a unitary weight because the branching process
is absent, similarly to the PDMC method the time step has to be chosen
short enough to produce only a small time step bias. 
For these two systems we found the time step of 0.01 hartree$^{-1}$
to be adequate to compute statistically exact results.

As a check of the ability of the PT-VMC method to compute
polarizabilities also for highly polarizable systems
whose exact wave function is more
diffuse than the one of He and H, we selected the hydride ion as test case.
For this system we optimized a 5 term Hylleraas-type wave function
whose average properties are shown in Table \ref{tab2} together
with the accurate results obtained in Ref. ~\cite{koga}.
Table \ref{tab2} contains also the multipole static polarizabilities computed
in this work employing a time step of 0.01 hartree$^{-1}$, 
and the static polarizability computed 
by Glover and Weinhold ~\cite{glo}.
Comparing the mean values in Table \ref{tab2}, one can notice that our 5 
term wave function gives lower values than the ones obtained in Ref.
~\cite{koga} except for $\left<r_-\right>$. 
This fact may explain the underestimation of the $\alpha_{dip}$
by PT-VMC, that recovers 92(2) percent of the accurate value.
Nevertheless, this result represents a fairly good estimation of the
static dipole polarizability for H$^-$, a quantity that appears difficult
to compute even with more complex approaches ~\cite{meyer}.

As far as PsH is concerned, we computed the autocorrelation functions
using two different trial wave functions, including one ($\Psi_T^1$)
and 28 ($\Psi_T^{28}$) terms ~\cite{maxcor}. 
The choice of two trial wave functions
to guide the Langevin dynamics was aimed at testing the dependency of
$C_{VV}(t)$ on the quality of the wave function itself. 

Employing the PT-VMC method and our wave functions for PsH, we computed
the autocorrelation functions for three perturbation potentials:

\begin{eqnarray}
V_1 = x_1 + x_2 - x_p \\
V_2 = \frac{3(x_1^2 +x_2^2 - x_p^2)-(r_1^2 +r_2^2 - r_p^2)}{2}\\
V_3 = x_1^3 +x_2^3 - x_p^3 -
\frac{3[x_1(y_1^2 + z_1^2)+x_2(y_2^2 + z_2^2)-x_p(y_p^2 + z_p^2)]}{2}
\label{pereq3}
\end{eqnarray}
\noindent
where the subscripts 1 and 2 indicate the two electrons, while the 
subscript $p$ indicates the positron.
These potentials are the cartesian forms of the dipole, quadrupole, and
octupole moment operators for the PsH system.
Figure 1 shows the averaged correlation functions for $V_1$, $V_2$, and $V_3$ as
obtained by the VMC method employing the 28 term trial wave function.
Each value of the correlation functions was computed employing roughly
10$^{10}$ configurations.
From Figure 1 one can note the effect at large evolution times
of the dispersion  of the
``trajectories'' used to compute the autocorrelation function.
This effect makes difficult the reproduction of the long-time regime
of these functions due to the exponential decay and the roughly
constant statistical error introduced by the method.
Moreover, the statistical error strongly depends on the perturbation
potential whose autocorrelation function is computed, i.e., more specifically
on the dispersion of its mean value over the $\Psi_T^2$ distribution.

The results for the static multipole polarizabilities, 
i.e. for $\omega = 0$,
computed with both trial wave functions, are shown in Table \ref{tab5}.
While for the dipole polarizabilities there is a good agreement between
the two values, larger differences are present for
the higher multipole polarizabilities. This fact is an indication
of the different accuracy of the two functions in approximating the exact
wave function at large distances from the nucleus. In fact it can be shown
that if one approximates the autocorrelation functions taking care
only of the excitation to the first state of the appropriate symmetry, 
the autocorrelation function is proportional to 
$\langle V_i^2 \rangle - \langle V_i \rangle^2$, where
$V_i$ is the perturbing potential.
Comparing the dipole results with the value obtained by Le Sech and Silvi ~\cite{sil},
again the large difference of the computed polarizabilities is apparent.
As a final check for this problem, we computed the energy of the
PsH when immersed in a weak static electric field $F$ 
by means of standard DMC simulations adding the linear potential 
$F(x_1 + x_2 - x_p)$. 
To make our simulations stable, i.e. to avoid the 
dissociation of the PsH, we truncated the effect of the linear potential
at $\left| x_i \right| = 15$ bohr. 
We fitted the DMC results by means of the simple
polynomial $a - \alpha_{dip} F^2 /2$, where $\alpha_{dip}$ is the static dipole
polarizability, obtaining $\alpha_{dip}=$ 42.3(8) a.u. 
We believe that this result, statistically indistinguishable from 
the $\alpha_{dip}$ obtained by the PT-VMC method, gives the definitive
answer to the problem of the PsH polarizability.
Nevertheless, the discrepancy between our PT-VMC and
DMC $\alpha_{dip}$ and the one computed by Le Sech and Silvi ~\cite{sil} 
remains puzzling. 
In our experience ~\cite{maxcor}, to compute the matrix elements they needed,
millions of configurations must be used even for systems like PsH
to avoid to be fooled by a false convergence.
Unfortunately, Le Sech and Silvi did not report any information about
the number of configurations they used to compute the integrals, so we
cannot judge the numerical accuracy of their results.

An attempt to estimate the total accuracy of our $\alpha$ results
can be made comparing the polarizability values obtained by the two
wave functions. These differ by 10 percent at most, a value that we feel
might give a conservative estimate of the relative errors for the higher 
multipolar fields.

As stated previously, although dynamical polarizabilities are 
interesting on their own, they represent the basis to compute van der
Waals dispersion coefficients for the interaction between different systems.
Therefore, following Ref. ~\cite{drake},
we present the calculation of the 
$C_6$, $C_8$, and $C_{10}$ dispersion coefficients
between  H, He, and PsH as a first effort to obtain accurate information
on the interaction between positronic systems and ordinary matter in the
framework of the Born-Oppenheimer approximation and second-order perturbation
theory.

Using the fitted parameters for H, He, and PsH we computed the coefficients 
for the interaction between the ordinary systems and between these and PsH. 
The values are reported in Table \ref{tab7}.
Since the values for the H-H, H-He, and He-He
coefficients are accurately known ~\cite{drake}, we use them as a test
of the accuracy of our approach: all the values differ from the accurate
results by Yan {\it et al.} ~\cite{drake} at most by one part over hundreds.

Comparing the $C_n$'s for the ordinary systems with the ones for the interaction
with PsH, it strikes that these last are more than an order of magnitude
larger than the formers. These features, due to the larger PsH polarizability,
indicates that positronic systems strongly interact with
ordinary matter even at large distances. 
Unfortunately, nothing can be said about location and depth
of the total potential minimum. This strongly depends also on the effect
of the repulsion between the positron cloud and the H and He nuclei, so that
we believe a supermolecule approach is needed.
In a previous work ~\cite{maxh2ps} we computed the interaction energy
between H and PsH, showing that this system could have a metastable state.
Although the dispersion coefficients for the interaction between He and PsH 
are smaller than those for PsH and H, they might be large enough to give rise
to a potential well that could support at least a stable state. If this
turns out to be the case, the He-PsH system could be the lightest 
van der Waals (i.e. bound by means of dispersion forces) stable dimer.

\heading{ACKNOWLEDGMENTS}
Financial support by the Universita' degli Studi di Milano 
is gratefully acknowledged. 
The authors are indebted to the
Centro CNR per lo Studio delle Relazioni
tra Struttura e Reattivita' Chimica for grants of computer time.

\pagebreak

{\bf Figure captions:}

\noindent
Figure 1. Logarithm of the correlation functions of the perturbing potentials.

\clearpage

\begin{table}
\begin{center}

\begin{tabular}{lcc}  \hline\hline
  & VMC$^a$ & Hylleraas$^b$ \\ \hline

$\langle E \rangle$& -0.52701(2)   & -0.52775$^b$    \\
$\langle V \rangle$& -1.0448(2)    & -1.0555$^b$    \\
$\langle r_{-} \rangle $& 2.7262   & 2.7102$^b$     \\
$\langle r_{-}^2 \rangle $& 11.844 & 11.915$^b$     \\
$\langle r_{--} \rangle $&4.4119   & 4.4127$^b$   \\
$\langle r_{--}^2 \rangle $&24.957 & 25.20$^b$    \\
$\alpha _{dip}$            &189.30 & 206(3)$^c$   \\
$\alpha _{quad}$           &5761.5 &              \\
$\alpha _{oct}$            &450758 &              \\
\hline

\end{tabular}
\caption{ Mean values for observables of the ground state $^1S$ of
H$^-$. All values are in atomic units.}
\label{tab2}
\end{center}
\end{table}

\noindent
$^a$ This work (5 term wave function) \\
$^b$ Ref. ~\cite{koga}\\
$^c$ Ref. ~\cite{glo}\\

\clearpage

\begin{table}
\begin{center}

\begin{tabular}{lccc}  \hline\hline
$\Psi_T$ & $\alpha_{dip}$ & $\alpha _{quad}$ & $\alpha_{oct}$ \\ \hline
$\Psi_T^{1}$ &  43.66(3)   & 972.7(2)   & 39178(32)  \\
$\Psi_T^{28}$ & 42.99(4) & 876.9(3)  & 34848(71)  \\

\hline \hline

\end{tabular}
\caption{Static multipole polarizabilities for the ground state $^{2,1}S$
of the PsH computed with one term ($\Psi_T^{1}$) and 28 term 
($\Psi_T^{28}$) wave functions. All values are in atomic units.}
\label{tab5}
\end{center}
\end{table}

\clearpage
\begin{table}
\begin{center}

\begin{tabular}{lccc}  \hline\hline
       &$C_6$  & $C_8$  &$C_{10}$ \\ \hline
H-H    & 6.480 & 125.23 & 3318.2  \\
       & 6.499$^a$ & 124.39$^a$ & 3285.8$^a$  \\
H-He   & 2.813 & 41.671 & 866.33  \\
       & 2.821$^a$ & 41.836$^a$ & 871.54$^a$  \\
He-He  &1.454  & 13.880 & 177.01  \\
       &1.461$^a$  & 14.117$^a$ & 183.69$^a$  \\
H-PsH  &40.30  & 2596.1 & 86292   \\
He-PsH &15.718 &950.80  & 23490   \\
\hline \hline

\end{tabular}
\caption{Computed dispersion coefficients. All values are in atomic units.}
\label{tab7}
\end{center}
\end{table}

\noindent
$^a$ Ref. ~\cite{drake}

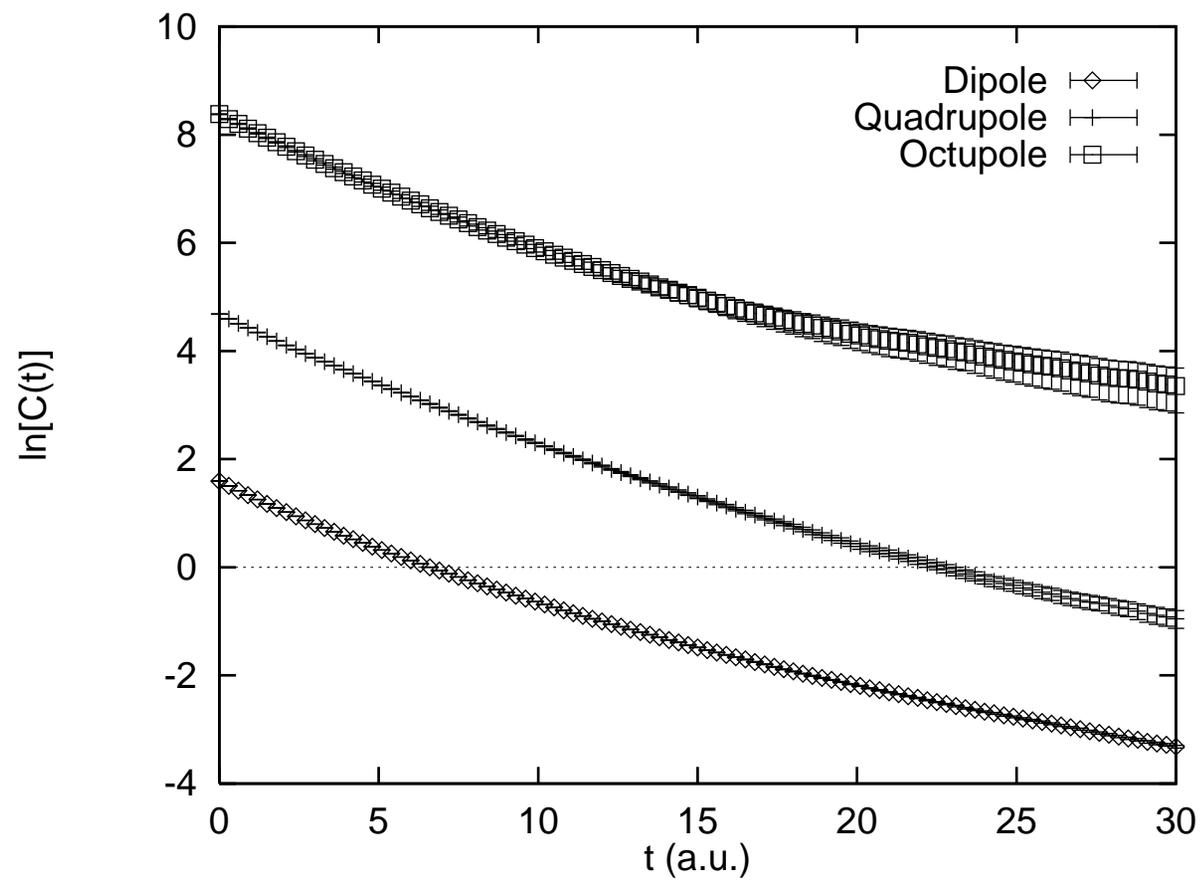
\end{document}